%% file: EMRecoilResub.tex
\documentclass[reprint,longbibliography,aip]{revtex4-2}

\usepackage{graphicx} 
\usepackage{amsmath}
\usepackage{ amssymb }

\usepackage{color}

\usepackage{float} 
\usepackage{wrapfig} 

\usepackage[makeroom]{cancel} 
\usepackage{comment}

\usepackage{lipsum} 

\usepackage{outlines}

\graphicspath{{./figs/}} 

\newcommand{\beq}{\begin{equation}}
\newcommand{\eeq}{\end{equation}}

\newcommand{\bvec}{\begin{pmatrix}}
\newcommand{\evec}{\end{pmatrix}}
\newcommand{\lp}{\left(}
\newcommand{\rp}{\right)}
\newcommand{\pa}[2]{\frac{\partial #1}{\partial #2}}

\newcommand{\tot}[2]{\frac{d #1}{d #2}}
\newcommand{\paf}[2]{\partial #1 / \partial #2}
\newcommand{\llangle}{\left \langle}
\newcommand{\rrangle}{\right \rangle}

\newcommand{\ve}[1]{\mathbf{#1}}
\newcommand{\kv}{\ve{k}}
\newcommand{\kapv}{\boldsymbol{\kappa}}

\newcommand{\Piv}{\boldsymbol{\Pi}}
\newcommand{\chiv}{\boldsymbol{\chi}}
\newcommand{\epsv}{\boldsymbol{\epsilon}}
\newcommand{\Ev}{\ve{E}}
\newcommand{\Bv}{\ve{B}}

\renewcommand{\Im}[1]{\text{Im} #1}
\renewcommand{\Re}[1]{\text{Re} #1}

\AtBeginDocument{%
\newwrite\bibnotes
\def\bibnotesext{Notes.bib}
\immediate\openout\bibnotes=\jobname\bibnotesext
\immediate\write\bibnotes{@CONTROL{REVTEX42Control}}
\immediate\write\bibnotes{@CONTROL{%
		aip42Control,author="08",editor="1",pages="1",title="0",year="1"}}
\if@filesw
\immediate\write\@auxout{\string\citation{aip42Control}}%
\fi
}%

\begin{document}



\title{Ponderomotive Recoil for Electromagnetic Waves}

\author{Ian E. Ochs}
\email{iochs@princeton.edu}
\affiliation{Department of Astrophysical Sciences, Princeton University, Princeton, New Jersey 08540, USA}
\author{Nathaniel J. Fisch}
\affiliation{Department of Astrophysical Sciences, Princeton University, Princeton, New Jersey 08540, USA}

\date{\today}

\begin{abstract}
	
When waves damp or amplify on resonant particles in a plasma, the nonresonant particles experience a recoil force that conserves the total momentum between the particles and electromagnetic fields. This force is important to understand, as it can completely negate current drive and rotation drive mechanisms that are predicted on the basis of only the resonant particles. Here, the existing electrostatic theory of this recoil force is extended to electromagnetic waves. While the result bears close similarity to historical fluid theories of laser-plasma interactions, it now incorporates both resonant and nonresonant particles, allowing momentum conservation to be self-consistently proven. Furthermore, the result is shown to be generally valid for kinetic plasmas, which is verified through single-particle hot-plasma simulations. The new form of the force provides physical insight into the nature of the generalized Minkowski (plasmon) momentum of geometrical optics, which is shown to correspond to the momentum gained by the field and nonresonant particles as the wave is self-consistently ramped up from vanishing amplitude.
	
\end{abstract}

\maketitle

\section{Introduction}

When a wave in a plasma interacts resonantly with a charged particle, the two can exchange substantial energy and momentum. If the absorbed momentum is parallel to the background magnetic field, current can be driven \cite{fisch1987theory}, while if it is perpendicular to the field, ExB rotation can be driven \cite{fetterman2008alpha}. 

However, the wave grows or damps due to the resonant interaction, meaning that its amplitude must change in time or space. These changing amplitudes introduce ponderomotive forces on the nonresonant particles—forces which can have an equally significant impact on the current and rotation drive. For instance, an electrostatic wave can grow as a result of interaction with a radial gradient of fusion-born fast alpha particles in a process known as alpha channeling. 
In the process, alpha particles diffuse radially outward across field lines via either Landau\cite{Fisch1992,fisch1992current,Heikkinen1996,ochs2015coupling,ochs2015alpha} or cyclotron\cite{Valeo1994,fisch1995ibw,Fisch1995a,Heikkinen1995,Herrmann1997,Marchenko1998,Kuley2011,Sasaki2011,Chen2016a,Gorelenkov2016,Cook2017,Castaldo2019,Cianfrani2018,Cianfrani2019,Romanelli2020,White2021AlphaParticle} resonances, and so one might expect a radial potential to develop, driving rotation. 
In a steady state problem, where a spatially-structured wave neither grows nor decays, the rotation effect survives \cite{Ochs2021WaveDriven,Ochs2022MomentumConservation}. 
However, as was recently shown, the initial value problem, where a plane wave grows in time, behaves very differently.
There, a self-consistent treatment shows that the charge transport from the outward diffusion of resonant alpha particles is completely canceled by an inward ponderomotive drift in the nonresonant bulk ions, completely eliminating the rotation drive effect\cite{ochs2021nonresonant}.

The ponderomotive response of the nonresonant particles is fundamentally related to momentum conservation. Resonant particles absorb the generalized Minkowski momentum $\ve{p}_M \equiv \ve{k} \mathcal{I}$ in the wave, where $\ve{k}$ is the wavevector and $\mathcal{I}$ is the wave action. However, the total plasma can only absorb the electromagnetic momentum in the wave, given by the Poynting flux $\ve{p}_{EM} = \llangle \Ev \times \Bv \rrangle / 4\pi c$, which vanishes for an electrostatic wave. Thus, the ponderomotive forces from a time-amplifying wave can be thought of as a nonresonant “recoil”, analogous to when a heavy cannon accelerates backward upon firing a cannonball. 

Previously, this momentum-conserving recoil has been derived for the case of a general electrostatic wave\cite{Ochs2020,ochs2021nonresonant,Ochs2021WaveDriven,Ochs2022MomentumConservation,Ochs2022Thesis}. Here, we extend this theory to the case of an electromagnetic wave. The end result is an expression which closely matches the result of past fluid theories, however now incorporating the resonant particles and derived in kinetic generality. Furthermore, the use of modern (Stix-like) wave formalism, separating the susceptibility tensor into its Hermitian and anti-Hermitian parts and Taylor expanding in small dissipation, allows momentum conservation to be shown straightforwardly. 

The electromagnetic approach used here also provides additional insight into the nature of the generalized Minkowski momentum itself. Typically, the Minkowski momentum is a property of a wave packet, which in a homogeneous plasma corresponds to the amount of momentum that must be taken from resonant particles in order to grow the wave from zero amplitude. Here, we explicitly show that this momentum is identical to the momentum that is added to the nonresonant particles and electromagnetic field (Poynting flux) as the wave grows. Thus, the Minkowski momentum, which is often treated abstractly in terms of the wave action, has two very intuitive physical meanings: it is simultaneously the amount of momentum (a) lost by the resonant particles and (b) gained by the nonresonant particles and electromagnetic field as the wave ramps up. 

The paper is organized as follows.
In Section~\ref{sec:derivation}, we derive the ponderomotive force on the plasma, first allowing for both spatial and temporal variation before focusing in on the initial value problem.
In so doing, we show that the force on the nonresonant particles of species $s$ can be written as a total time derivative of a simple nonresonant momentum $\ve{p}_{Ns}$.
In Section~\ref{sec:conservation}, we make use of the dispersion relation to show that the resulting theory conserves momentum between resonant particles, nonresonant particles, and electromagnetic fields.
In Section~\ref{sec:minkowski}, we futher show that the generalized Minkowski momentum of geometrical optics represents the sum of the nonresonant momentum $\ve{p}_{Ns}$ and electromagnetic momentum $\ve{p}_{EM}$.

In the remaining sections, we focus on validating our results and comparing to the existing literature.
In Section~\ref{sec:fluidComparison}, we compare our results to the existing fluid theories of the time-dependent ponderomotive force, showing agreement.
To validate that these expressions truly work for kinetic plasmas, in Section~\ref{sec:simulations} we perform single-particle simulations of a ramping electromagnetic wave interacting with hot particles in a magnetized plasma.
We show that the resulting momentum can be derived from the standard hot plasma susceptibility, confirming the validity of the theory.


\section{Derivation of the Nonresonant Recoil Momentum} \label{sec:derivation}

In this section, we explicitly calculate the force on a plasma species from the electromagnetic fields of a quasi-monochromatic wave, in terms of the species contribution to the plasma susceptibility.
For simplicity, we will assume the plasma is homogeneous and stationary, to allow us to robustly use Fourier methods rather than Weyl methods\cite{Dodin2022QuasilinearTheory}.

Our starting point is the Lorentz force on the local Eulerian volume:
\begin{align}
	\ve{F}_s &= \rho_s \ve{E} + \frac{1}{c} \ve{j}_s \times \ve{B}.
\end{align}
Here, $\Ev$ and $\Bv$ are the electric and magnetic fields, and $\rho_s$ and $\ve{j}_s$ are the charge and current densities of species $s$.

If we assume that all quantities $A \in \{\rho,\ve{j},\Ev,\Bv\}$ vary as $A = \tilde{A}_0 e^{i\ve{k} \cdot \ve{x} - i \omega t}$, then by averaging over a cycle, we find:
\begin{align}
	\llangle \ve{F}_s\rrangle &= \frac{1}{2} \Re \left[ \tilde{\rho}_{s0} \ve{\tilde{E}}_0^* + \frac{1}{c} \ve{\tilde{j}}_{s0} \times \ve{\tilde{B}}_0^* \right] e^{-2 \kapv \cdot \ve{x} +2 \omega_i t},
\end{align}
where $\kapv \equiv \Im (\ve{k})$ and $\omega_k \equiv \Im (\omega)$.
Defining $\tilde{A} = \tilde{A}_0 e^{- \kapv \cdot \ve{x} + \omega_i t}$ to encode the local wave amplitude, we can write this force as:
\begin{align}
	\llangle \ve{F}_s\rrangle &= \frac{1}{2} \Re \left[ \tilde{\rho}_s \ve{\tilde{E}}^* + \frac{1}{c} \ve{\tilde{j}}_s \times \ve{\tilde{B}}^* \right]. \label{eq:FourierLorentz}
\end{align}

Now we can begin to write all these quantities in terms of the electric field.
The magnetic field is most straightforward; from Faraday's Law:
\begin{align}
	\nabla \times \ve{E} &= -\frac{1}{c} \pa{\ve{B}}{t}.
\end{align}
After Fourier transformation, this becomes:
\begin{align}
	\ve{\tilde{B}} &= \frac{c}{\omega} \ve{k} \times \ve{\tilde{E}}. \label{eq:BERelation}
\end{align}

To express the currents and charge densities, we must make use of the plasma susceptibility.
In a homogeneous stationary medium, the current density can be written for some susceptibility $\chiv_s$ as \cite{stix1992waves}:
\begin{align}
	\ve{\tilde{j}}_s &= -\frac{i}{4\pi} \omega \chiv_s \cdot \ve{\tilde{E}}. \label{eq:j}
\end{align}

From this current density, we can get the charge density via the continuity equation:
\begin{align}
	\pa{\rho_s}{t} = -\nabla \cdot \ve{j}_s.
\end{align}
After Fourier transforming, we find:
\begin{align}
	\tilde{\rho}_s = \frac{\ve{k} \cdot \ve{\tilde{j}}_s}{\omega} = -\frac{i}{4\pi} \ve{k} \cdot \chiv_s \cdot \ve{\tilde{E}}. \label{eq:rho}
\end{align}

From this point on, we drop the tildes on the electric and magnetic fields, since we work exclusively in Fourier space.
Plugging Eqs.~(\ref{eq:BERelation}), (\ref{eq:j}), and (\ref{eq:rho}) into Eq.~(\ref{eq:FourierLorentz}), we find:
\begin{align}
	\llangle \ve{F}_s\rrangle &= \frac{1}{2} \Re \biggl[ \lp -\frac{i}{4\pi} \ve{k} \cdot \chiv_s \cdot \ve{E} \rp  \ve{E}^* \notag\\
	&\hspace{0.5in}  + \frac{1}{c} \lp -\frac{i}{4\pi} \omega \chiv_s \cdot \ve{E} \rp \times \lp \frac{c}{\omega^*} \ve{k}^* \times \ve{E}^* \rp \biggr].\\
	&= \frac{1}{8\pi} \Im \biggl\{ \lp \ve{k} \cdot \chiv_s \cdot \ve{E} \rp  \ve{E}^* + \frac{\omega}{\omega^*} \left[ \lp \chiv_s \cdot \ve{E} \rp \times \lp \kv^* \times \Ev^*\rp \right] \biggr\}. \label{eq:Fs1}
\end{align}

Thus far, our expressions have been true for arbitrary complex $\omega$ and $\ve{k}$.
However, our focus on quasi-monochromatic waves allows us to proceed further, by expanding in small $\omega_i/\omega_r$ and $|\kapv|/|\kv|$.
Working to first order, we can rewrite the first term in brackets on the RHS:
\begin{align}
	A &\equiv \frac{\omega}{\omega^*} \left[ \lp \chiv_s \cdot \ve{E} \rp \times \lp \kv^* \times \Ev^*\rp \right]\\
	&\approx \left[ \lp \ve{E}^* \cdot \chiv_s \cdot \ve{E} \rp \ve{k}^* - \lp \ve{k}^* \cdot \chiv_s \cdot \ve{E} \rp \ve{E}^* \right] \notag\\
	&\quad  + 2 i\frac{\omega_i}{\omega_r} \lp \chiv_s \cdot \ve{E} \rp \times \lp \kv_r \times \Ev^*\rp, \label{eq:intStep1}
\end{align}
where in the first line we used the double cross product identity, and in the second we ignored terms of order $\kappa \omega_i / k \omega_r$.
Now, the second term in brackets in Eq.~(\ref{eq:intStep1}) combines with the first term in braces in Eq.~(\ref{eq:Fs1}), together becoming $2 i\lp \kapv \cdot \chiv_s \cdot \ve{E} \rp  \ve{E}^*$.
This leaves only the first term in brackets left to simplify:
\begin{align}
	\Im \left[ \lp \ve{E}^* \cdot \chiv_s \cdot \ve{E} \rp \ve{k}^* \right] &= \Im [ \lp \ve{E}^* \cdot (\chiv_s)_H \cdot \ve{E} \rp (-i \kapv_r) \notag \\
	&\qquad + i \lp \ve{E}^* \cdot (\chiv_s)_A \cdot \ve{E} \rp i \kv_r ].
\end{align}
Here, $(\chiv_s)_H$ and $(\chiv_s)_A$ are the Hermitian and anti-Hermitian parts of $\chi$, when evaluated at complex $\omega$ and $\ve{k}$.
Denoting $\chiv_s^H$ and $\chiv_s^A$ as the Hermitian and anti-Hermitian parts of $\chiv_s$ at real $\omega$ and $\ve{k}$, and assuming that $|\chiv_s^H| \gg |\chiv_s^A|$, we can Taylor expand near the real frequencies to find:
\begin{align}
	(\chiv_s)_H &\approx \chiv_s^H\\
	(\chiv_s)_A &\approx \chiv_s^A + \omega_i \pa{\chiv_s^H}{\omega_r} + \kapv \cdot \pa{\chiv_s^H}{\kv_r}.
\end{align}

Putting this all together and plugging back in to Eq.~(\ref{eq:Fs1}), we find:
\begin{align}
	\llangle \ve{F}_s\rrangle &= \frac{1}{4\pi} \Re \biggl\{ \ve{E}^* \lp \kapv \cdot \chiv_s^H \cdot \ve{E} \rp    - \frac{1}{2} \kapv \lp \Ev^* \cdot \chiv_s^H \cdot \Ev \rp \notag\\
	&\hspace{0.1in} + \omega_i\left[ \lp \chiv_s^H \cdot \ve{E} \rp \times \lp \frac{\kv_r}{\omega_r} \times \Ev^*\rp \right]  \notag\\
	&\hspace{0.1in} + \frac{1}{2}\ve{k}_r \lp \ve{E}^* \cdot \lp \chiv_s^A + \omega_i \pa{\chiv_s^H}{\omega_r} + \lp \kapv \cdot \pa{}{\ve{k}_r} \rp \chiv_s^H \rp  \cdot \ve{E} \rp \biggr\}. \label{eq:FsFull}
\end{align}
As shown in Appendix~\ref{app:EMES}, this force is the electromagnetic generalization of the electrostatic force first derived by Kato\cite{kato1980electrostatic} for parallel forces using the magnetized kinetic dispersion relation, and later generalized to any electrostatic wave \cite{Ochs2020,Ochs2021WaveDriven}.

Additional insight can be gained by recasting the force in terms of the temporal and spatial derivatives of the wave field.
Recalling that $\Ev, \Ev^* \sim e^{-\kapv \cdot \ve{x} + \omega_i t}$, we note that in Eq.~\ref{eq:FsFull} we can make the substitutions $2\omega_i \rightarrow \paf{}{t}$ and $2\kapv  \rightarrow -\paf{}{\ve{x}}$.
Thus, we can write the force as the sum of (a) the divergence of a nonresonant stress $\Piv_{Ns}$, (b) the time derivative of a nonresonant momentum $\ve{p}_{Ns}$ (which we term the \emph{nonresonant recoil} \cite{Ochs2020,ochs2021nonresonant,Ochs2021WaveDriven}), and (c) a resonant dissipation term $\ve{F}_{Rs}$:
\begin{align}
	\llangle \ve{F}_s\rrangle = -\pa{}{\ve{x}} \cdot \Piv_{Ns} + \pa{}{t} \ve{p}_{Ns} + \ve{F}_{Rs},
\end{align}
where
\begin{align}
	\Piv_{Ns} &= \frac{1}{8\pi} \Re \biggl[\lp \chiv_s^H \cdot \ve{E} \rp \ve{E}^*   -  \frac{1}{2} \ve{I} \lp \Ev^* \cdot \chiv_s^H \cdot \Ev \rp \notag\\
	&\hspace{0.6in} + \frac{1}{2} \lp \pa{}{\ve{k}_r} \lp \Ev^* \cdot \chiv_s^H \cdot \Ev \rp  \rp \ve{k}_r  \biggr]\\
	\ve{p}_{Ns} &= \frac{1}{8\pi} \Re \biggl[ \lp \chiv_s^H \cdot \ve{E} \rp \times \lp \frac{\kv_r}{\omega_r} \times \Ev^*\rp \notag\\
	&\hspace{0.6in} + \frac{\ve{k}_r}{2} \pa{}{\omega_r} \lp \Ev^* \cdot \chiv_s^H \cdot \Ev \rp \biggr]\\
	\ve{F}_{Rs} &= \frac{1}{8\pi} \ve{k}_r \lp \Ev^* \cdot \chiv_s^A \cdot \Ev \rp.
\end{align}

In this paper, we will focus primarily on plane waves which evolve temporally, and so we will largely ignore the stress term.
However, there are a couple points to note.
First, the stress term in this form represents the electromagnetic force on the plasma volume, and thus the sum of stress on all species is consistent with the Maxwell stress tensor (Appendix~\ref{app:MaxwellStress}).
Second, this form of the stress is of limited utility, since it must be combined with Reynolds and polarization stress terms \cite{Jaeger2000,Myra2000,Myra2004,Gao2007,Ochs2021WaveDriven,Ochs2022Thesis} in order to yield a meaningful total ponderomotive force.

Unlike the stress term, the resonant dissipation and nonresonant recoil terms do not suffer from interpretive difficulty.
As we show in the next section, together, they conserve momentum between the resonant particles, nonresonant particles, and electromagnetic field.

Before moving on, however, we note that the nonresonant momentum $\ve{p}_{Ns}$ can be rewritten in several ways, which are useful in different contexts depending on the wave polarization and structure of $\chiv_s$.
Defining the refractive index $\ve{n} = \ve{k} c / \omega$, we can write:
\begin{align}
	\ve{p}_{Ns} &= \frac{1}{8\pi c} \Re \biggl[ \lp \chiv_s^H \cdot \ve{E} \rp \times \lp \ve{n}_r \times \Ev^*\rp \notag\\
	&\hspace{0.6in} + \frac{\ve{n}_r}{2} \omega \pa{}{\omega_r} \lp \Ev^* \cdot \chiv_s^H \cdot \Ev \rp \biggr] \label{eq:pNsN}\\
	\ve{p}_{Ns}&= \frac{1}{8\pi c} \Re \biggl[ \ve{n}_r \lp \ve{E}^* \cdot \chiv_s^H \cdot \ve{E} \rp - \ve{E}^* \lp \ve{n}_r \cdot \chiv_s^H \cdot \ve{E} \rp  \notag\\
	&\hspace{0.6in} + \frac{\ve{n}_r}{2} \omega \pa{}{\omega_r} \lp \Ev^* \cdot \chiv_s^H \cdot \Ev \rp \biggr]\\
	\ve{p}_{Ns}&= \frac{1}{16\pi c} \Re \biggl[ \frac{\ve{n}_r}{\omega}  \pa{}{\omega_r} \lp \omega^2 \ve{E}^* \cdot \chiv_s^H \cdot \ve{E} \rp   \notag\\
	&\hspace{0.6in} - 2 \ve{E}^* \lp \ve{n}_r \cdot \chiv_s^H \cdot \ve{E} \rp \biggr] \label{eq:pNs3}.
\end{align}

\section{Momentum Conservation} \label{sec:conservation}

Having calculated the form of the forces on resonant and nonresonant particles, we show in this section that these forces respect momentum conservation.
We focus on the case of a plane wave which grows or damps only in time, neglecting the $\kapv$-dependent terms in the force.

To demonstrate momentum conservation, we will have to eliminate the susceptibilities $\chiv_s$.
To do this, we make use of the general dispersion relation for an electromagnetic wave \cite{stix1992waves}:
\begin{align}
	\ve{n} \times \lp \ve{n} \times \ve{E} \rp + \lp \ve{I} + \sum_s \chiv_s \rp \cdot \ve{E} = 0. \label{eq:EMDispersion}
\end{align}
From this dispersion relation, we can derive two useful identities.

First, to zeroth order in $\epsilon \sim |\chiv_s^A| / |\chiv_s^H| \sim |\omega_i| / |\omega_r|$, we have:
\begin{align}
	\sum_s \chiv_s^H \cdot \ve{E} &= -\ve{E} - \ve{n}_r \times \lp \ve{n}_r \times \ve{E} \rp + \mathcal{O}(\epsilon) \label{eq:DispZerothOrder}.
\end{align}

Second, by dotting in $\ve{E}^*$, we find to first order in $\epsilon$: 
\begin{align}
	\ve{E}^* \cdot \sum_s (\chiv_s)_A \cdot \ve{E} &= - \ve{E}^* \cdot \ve{n}_i \times \lp \ve{n}_r \times \ve{E} \rp \notag\\
	& \quad - \ve{E}^* \cdot \ve{n}_r \times \lp \ve{n}_i \times \ve{E} \rp\\
	&= 2\frac{\omega_i}{\omega_r} \ve{E}^* \cdot \ve{n}_r \times \lp \ve{n}_r \times \ve{E} \rp + \mathcal{O}(\epsilon^2)
\end{align}
In the last line, we used the fact that $\ve{a} \times \lp \ve{a} \times \ve{E} \rp$ is a Hermitian operator for real $\ve{a}$, along with the Taylor expansion:
\begin{align}
	\ve{n}_i &= \Im \lp \frac{\ve{k} c}{(\omega_r + i\omega_i)} \rp \approx -\frac{\omega_i}{\omega_r} \ve{n}_r.
\end{align}

With these two identities, we are in a position to prove momentum conservation to the relevant first order in $\epsilon$.
Taking $\kapv = 0$ and summing Eq.~(\ref{eq:FsFull}) over all species, we have:
\begin{align}
	\sum_s \llangle \ve{F}_s\rrangle &= \frac{1}{4\pi} \Re \biggl\{ \omega_i \lp -\ve{E} - \ve{n}_r \times \lp \ve{n}_r \times \ve{E} \rp \rp \times \lp \frac{\kv_r}{\omega_r} \times \Ev^*\rp   \notag\\
	&\hspace{0.1in} + \frac{1}{2}\ve{k}_r \lp  2\frac{\omega_i}{\omega_r} \ve{E}^* \cdot \ve{n}_r \times \lp \ve{n}_r \times \ve{E} \rp \rp \biggr\}\\
	&= -\frac{\omega_i}{4 \pi c} \Re \lp \ve{E} \times \ve{B}^*  + \ve{C} \rp,
\end{align}
where 
\begin{align}
	\ve{C} &\equiv \Re \biggl\{\left[ \ve{n}_r \times \lp \ve{n}_r \times \ve{E} \rp \right]  \times \lp \ve{n}_r \times \Ev^*\rp \notag \\
	&\hspace{0.55in}-\ve{n}_r \left[  \ve{E}^* \cdot \ve{n}_r \times \lp \ve{n}_r \times \ve{E} \rp  \right] \biggr\}\\
	&= \Re \biggl\{\cancel{\left[  \ve{n}_r \cdot \lp \ve{n}_r \times \Ev^*\rp \right] } \lp \ve{n}_r \times \ve{E} \rp  \notag \\
	&\hspace{0.55in} - \ve{n}_r  \left[ \lp \ve{n}_r \times \ve{E} \rp \cdot \lp \ve{n}_r \times \Ev^*\rp \right] \notag\\
	&\hspace{0.55in}+\ve{n}_r \left[  \lp \ve{n}_r \times \ve{E}^* \rp \cdot  \lp \ve{n}_r \times \ve{E} \rp  \right]\biggr\}\\
	&= 0.
\end{align}

Thus, recalling that we can identify $2 \omega_i \rightarrow \paf{}{t}$, and recalling that the time average of two oscillating quantities $\Re(\ve{E} \times \ve{B}^*) \rightarrow 2 \llangle \ve{E} \times \ve{B} \rrangle$, we have:
\begin{align}
	\sum_s \llangle \ve{F}_s\rrangle &= -\pa{}{t} \lp \frac{\llangle \ve{E} \times \ve{B}  \rrangle}{4 \pi c}  \rp = -\pa{\ve{p}_{EM}}{t},
\end{align}
where $\ve{p}_{EM} = \ve{S}_{EM}/c^2$ is the electromagnetic momentum.
Thus, the momentum gained by the resonant and nonresonant particles is precisely the momentum lost by the electromagnetic fields.

\section{Relationship to Generalized Minkowski Momentum} \label{sec:minkowski}

Having demonstrated momentum conservation between waves and particles, we can gain additional insight by looking at different combinations of the momenta. 
To this end, begin by summing up the nonresonant momenta of each species (Eq.~(\ref{eq:pNsN})), and then plug in the zeroth-order dispersion relation from Eq.~(\ref{eq:DispZerothOrder}) and the relationship between $\ve{B}$ and $\ve{E}$ fields from Eq.~(\ref{eq:BERelation}).
After a few vector manipulations, including noting that $\ve{n} \cdot \ve{B} = \ve{n} \cdot (\ve{n} \times \Ev) = 0$, we find to lowest order:
\begin{align}
	\sum_s \ve{p}_{Ns} &= \frac{1}{8\pi c} \Re \biggl[- \Ev^* \times \Bv \notag\\
	&\quad + \ve{n} \lp \Bv^* \cdot \Bv \rp + \frac{\ve{n}}{2} \Ev^* \cdot \lp \omega \pa{\epsv}{\omega} \rp \cdot \Ev \biggr]
\end{align}

We can recognize the first term in brackets as the (negative) Poynting momentum flux.
Thus, the sum of electromagnetic and nonresonant momentum takes the form:
\begin{align}
	\ve{p}_{EM} + \sum_s \ve{p}_{Ns} &= \frac{\ve{n}}{8\pi c}  \biggl[ \Bv^* \cdot \Bv + \frac{1}{2} \Ev^* \cdot \lp \omega \pa{\epsv}{\omega} \rp \cdot \Ev \biggr]. \label{eq:pNonresExB}
\end{align}

This combination of the momentum is useful not only because it is compact, but also because it is familiar.
In the study of geometrical optics, the wave can be identified with a ``plasmon'' \cite{Tsytovich1977} or ``generalized Minkowski'' \cite{dodin2012axiomatic} momentum.
Derived from Noether's theorem for the quasi-monochromatic wave Lagrangian, this momentum appears in the conservation laws governing the evolution of the wave envelope in the presence of dissipation or refraction, and can be thought of as the canonical momentum of the wave photons.
It can be written as\cite{dodin2012axiomatic}:
\begin{align}
	\ve{p}_M &= \ve{k} \mathcal{I},
\end{align}
where $\mathcal{I}$ is the wave action, given for an electromagnetic wave in a dispersive dielectric by:
\begin{align}
	\mathcal{I} &= \frac{1}{16 \pi \omega} \left[ \Ev^* \cdot \pa{}{\omega}(\epsv \omega) \cdot \Ev + \Bv^* \cdot \Bv \right]. \label{eq:EMAction1}
\end{align}

This expression can be made more familiar by making use of the zeroth-order dispersion relation (Eq.~(\ref{eq:DispZerothOrder})) and the relationship between $\ve{B}$ and $\ve{E}$ fields from Eq.~(\ref{eq:BERelation}), which together imply that $\Ev^* \cdot \epsv  \cdot \Ev = \Bv^* \cdot \Bv$ to lowest order.
Application of the chain rule then to Eq.~(\ref{eq:EMAction1}) then quickly yields:
\begin{align}
	\ve{p}_M &= \frac{\ve{n}}{16 \pi c} \left[ \Ev^* \cdot \lp \omega \pa{\epsv}{\omega} \rp \cdot \Ev + \Ev^* \cdot \epsv  \cdot \Ev + \Bv^* \cdot \Bv \right]\\
	&= \frac{\ve{n}}{16\pi c}  \biggl[ \Ev^* \cdot \lp \omega \pa{\epsv}{\omega} \rp \cdot \Ev + 2\Bv^* \cdot \Bv\biggr].
\end{align}
We can see that this is the same quantity as in Eq.~(\ref{eq:pNonresExB}), so that:
\begin{align}
	\ve{p}_M &= \ve{p}_{EM} + \sum_s \ve{p}_{Ns}.
\end{align}
Thus, the Minkowski momentum, usually interpreted as the canonical momentum of a photon which governs the wave packet evolution, also has a quite intuitive physical interpretation.
Namely, it is the combined momentum gained by the electromagnetic wave field and the nonresonant particles as the wave grows from 0 amplitude.

\section{Relationship to Klima-Petrilzka Fluid Ponderomotive Force} \label{sec:fluidComparison}

The form of the ponderomotive force has been the focus of study for many years, with much of the work focused on transfer laser-driven implosions through Washimi-Karpan forces \cite{Washimi1976PonderomotiveForce}.
As reviewed extensively in Kentwell \cite{Kentwell1987}, many closely related fluid theories were developed to study these forces\cite{Karpman1982PonderomotiveForce,Vukovic1984PonderomotiveForce,Klima1978RadiationPressure,Lee1983PonderomotiveForce}.
These papers largely found the same expression for the time-dependent recoil force on nonresonant particles.
This is Eq.~(4.69) in Kentwell \cite{Kentwell1987} (though importantly the last term has a typo in the location of the parentheses, which should end before the fields), and Eq.~(30) in Lee and Parks \cite{Lee1983PonderomotiveForce}.

The fluid models were employed to simplify the calculation of the forces in the presence of spatial gradients, which require complicated evaluations of Reynolds stresses in the kinetic theory.
However, for the time-dependent recoil, nothing in our analysis above assumed that the plasma was describable by a fluid model. 
Therefore, this expression is in fact more general than previously thought.
To verify this generality, in the next section, we perform single-particle simulations showing that the recoil force matches the theoretical expression for the ponderomotive force even for highly kinetic plasmas.

%
%

\section{Kinetic Simulations} \label{sec:simulations}

In this section, we use single-particle simulations to verify that the theory of the nonresonant recoil works for kinetic plasmas.
For simplicity, we consider a magnetized plasma with $\ve{B}_0 \parallel \hat{z}$, and a wave with $\ve{k} \parallel \hat{z}$ and $\ve{E} \parallel \hat{x}$.
From Eq.~(\ref{eq:pNs3}), we see that this choice means that we only need the $\chi_{s,xx}$ component of the susceptibility tensor to calculate the force along the $\hat{z}$ direction.

To find $\chi_{s,xx}$, we can use the hot plasma susceptibility tensor from Stix \cite{stix1992waves} Section 10.6, given (dropping species subscripts $s$) by:
\begin{align}
	\chi_{xx} &= \frac{\omega_{p}^2}{\omega \Omega} \int_0^{\infty} 2 \pi v_\perp' dv_\perp' \int_{-\infty}^{\infty} dv_\parallel' \notag\\ &\quad \times \sum_{n=-\infty}^{\infty} \frac{\Omega v_\perp'}{\omega - k_\parallel v_\parallel' - n \Omega} \frac{n^2 J_n(z)^2}{z^2}. 
\end{align}
Here, $\omega_p = \sqrt{4\pi n_s q_s^2 / m_s}$ is the plasma frequency, $\Omega = q_s B/c m_s$ is the gyrofrequency, $q_s$, $m_s$, and $n_s$ are the charge, mass, and density of species $s$, $J_n(z)$ is the Bessel function of the first kind, and:  
\begin{align}
	U &\equiv \lp 1 - \frac{k_\parallel v_\parallel'}{\omega} \rp \pa{f_0}{v_\perp'} + \frac{k_\parallel v_\perp'}{\omega} \pa{f_0}{v_\parallel'}\\
	z &\equiv \frac{k_\perp v_\perp'}{\Omega}.
\end{align}

To check this kinetically using single-particle simulations, we will want to express this $\chi_{xx}$ for the distribution $f_0 = \delta(v_\perp' - v_\perp) \delta(v_\parallel' - v_\parallel) / 2 \pi v_\perp$.
Note that once the form of the force is verified for this distribution, it is verified for more general distributions with finite velocity spreads as well, since the total force can be composed as an integral of the $\delta$-function force over the velocity distribution.
This choice of $f_0$ yields, after integration by parts to eliminate the derivatives of $f$ contained in $U$:
\begin{align}
	\chi_{xx} &= -\frac{\omega_p^2}{\omega^2} \sum_n \biggl[ \lp \omega - k_\parallel v_\parallel \rp \frac{1}{v_\perp} \pa{Y_n}{v_\perp}  + k_\parallel \pa{Y_n}{v_\parallel} \biggr],\\
	 Y_n &\equiv \frac{v_\perp^2}{\omega-k_\parallel v_\parallel - n \Omega} \frac{n^2 J_n(z)^2}{z^2}.
\end{align}

Plugging this into Eq.~(\ref{eq:pNs3}) for $k_\perp = 0$ yields:
\begin{align}
	p_{Ns,z} = k_\parallel \frac{|E_x|^2}{8\pi} \frac{\omega_p^2}{\omega^2} \frac{\alpha \Omega^2}{(\alpha^2-\Omega^2)^2} \left[1 + \frac{1}{2} \frac{k_\parallel^2 v_\perp^2}{\Omega^2} \frac{\alpha^2 + 3 \Omega^2}{\alpha^2 - \Omega^2} \right],
\end{align}
where $\alpha = \omega - k_\parallel v_\parallel$ is the Doppler-shifted frequency of the wave.

Now, we can divide $p_{Ns,z}$ by the density to get the nonresonant momentum of a single particle. 
If we additionally divide by the mass, then we end up with the the final velocity of a particle if we ramp up a wave from 0 amplitude (assuming that the force is small enough that $v_\parallel$ can be treated as constant).
If we additionally write the electric field in terms of a vector potential:
\begin{align}
	\ve{E} &= -\frac{\omega}{c} \ve{A},
\end{align}
then we find:
\begin{align}
	\Delta v_z =  \frac{k_\parallel |A_x|^2 q^2}{2 m^2 c^2}  \frac{\alpha \Omega^2}{(\alpha^2-\Omega^2)^2} \left[1 + \frac{1}{2} \frac{k_\parallel^2 v_\perp^2}{\Omega^2} \frac{\alpha^2 + 3 \Omega^2}{\alpha^2 - \Omega^2} \right]. \label{eq:xPolDeltaVz}
\end{align}
Because the hot plasma susceptibility involves gyro-averaging the plasma response, this ponderomotive force corresponds to the force on a ring of charge with fixed initial $v_\perp$.
Thus, to check for agreement, Eq.~(\ref{eq:xPolDeltaVz}) must be compared to the average change in final velocity of an ensemble of particles with different initial gyro-angle, in the presence of a wave that ramps up slowly compared to the wave- and gyro-periods.

\begin{figure*}[t]
	\centering
	\includegraphics[width=\linewidth]{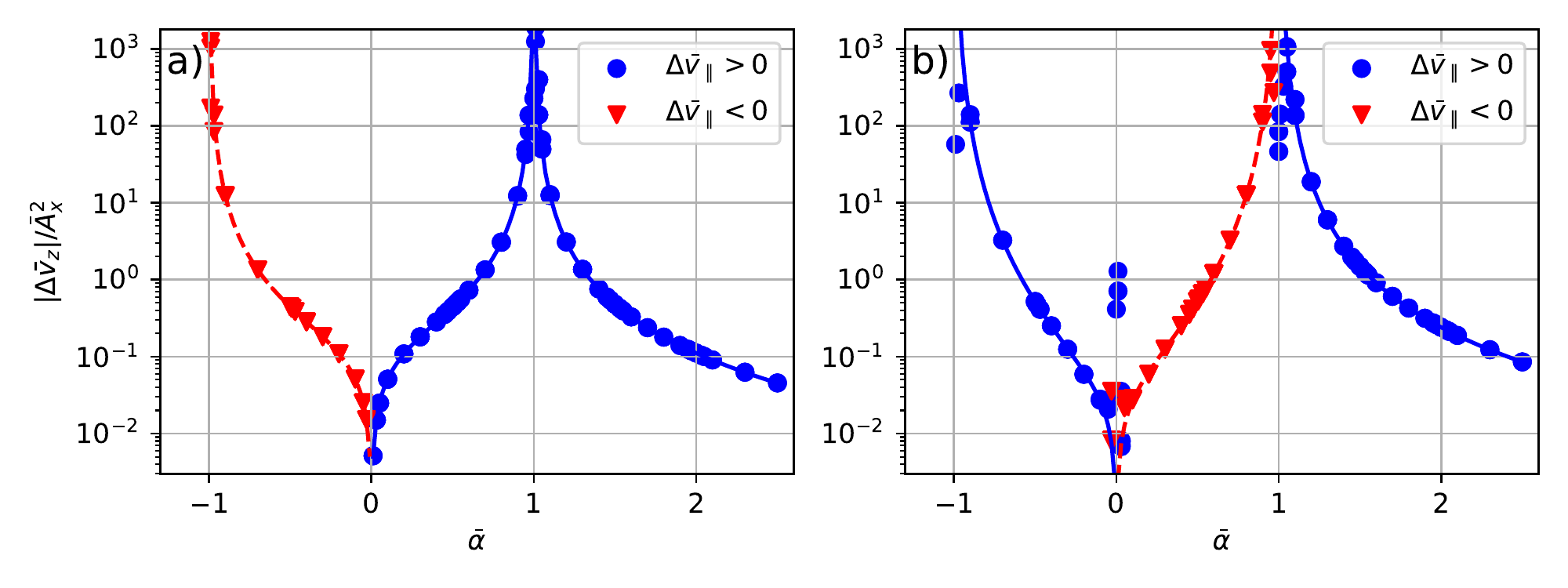}
	\caption{Change in normalized parallel velocity $\bar{v}_z = v_z \Omega/k_\parallel$ as a function of normalized Doppler-shifted initial frequency $\bar{\alpha} = (\omega - k_\parallel v_\parallel)/\Omega$ for a gyro-averaged ensemble of particles, as a result of ramping up an $\hat{x}$-polarized wave with $\ve{k} \parallel \ve{B}$ from 0 amplitude.
		Results are shown for (a) $\bar{v}_\perp = 0$, and (b) $\bar{v}_\perp = 1$ (i.e. $k_\parallel \rho = 1$).
		Theoretical predictions from Eq.~(\ref{eq:xPolDeltaVzBar}) are shown as lines, with solid blue lines for positive values and dashed red lines for negative values. 
		Simulations are shown as markers, with blue circles for positive values and red triangles for negative values.
		The agreement is quite good, except near the gyroresonances, where the nonresonant approximation breaks down.}
	\label{fig:xPolSim}
\end{figure*}

To perform this comparison to simulations, it helps to first nondimensionalize.
We take:
\begin{alignat}{3}
	&\bar{t} &&= \Omega t  \hspace{0.5in}  &&\bar{\ve{x}} = k_\parallel \ve{x} \\
	&\bar{\omega} &&= \omega/\Omega  &&\bar{\ve{v}}= k_\parallel \Omega^{-1} \ve{v} \\
	&\bar{\ve{A}} &&= k_\parallel B_0^{-1} \ve{A}.\hspace{0.25in} &&
\end{alignat}
In these units, the wave field is given by:
\begin{align}
	\bar{\ve{A}} &= -g(\bar{t}) \bar{A}_0 \cos \lp \bar{z} - \bar{\omega} \bar{t} \rp \hat{x} \\
	g(\bar{t}) &\equiv \min\lp\frac{\bar{t}}{\bar{\tau}_\text{R}},1\rp
\end{align}
where $\bar{\tau}_R \gg 2\pi/\min(\omega,\Omega)$ is the ramp time of the wave.
From Eq.~(\ref{eq:xPolDeltaVz}), the theoretical change in velocity after the wave rampup is then:
\begin{align}
	\frac{\Delta \bar{v}_z}{|\bar{A}_x|^2} =  \frac{1}{2}  \frac{\bar{\alpha} }{(\bar{\alpha}^2-1)^2} \left[1 + \frac{\bar{v}_\perp^2}{2} \frac{\bar{\alpha}^2 + 3 }{\bar{\alpha}^2 - 1} \right], \label{eq:xPolDeltaVzBar}
\end{align}
where $\bar{\alpha}$ is the normalized Doppler-shifted frequency:
\begin{align}
	\bar{\alpha} &\equiv \bar{\omega} - \bar{v}_\parallel = \frac{\omega - k_\parallel v_\parallel}{\Omega}.
\end{align}

The dimensionless Lorentz force to be simulated is given by:
\begin{align}
	\tot{\ve{\bar{v}}}{\bar{t}} &= -\pa{\bar{\ve{A}}}{\bar{t}} + \bar{\ve{v}} \times \lp \bar{\nabla} \times \bar{\ve{A}}  + \hat{z} \rp.
\end{align}
This equation of motion corresponds to a Boris-style push with $q/m = 1$, $\ve{E} = -\paf{\bar{\ve{A}}}{\bar{t}}$, and $\ve{B} = \bar{\nabla} \times \bar{\ve{A}}  + \hat{z}$.
To retain accuracy over very long timescales, we used a recently-developed modification to the Boris push algorithm \cite{Zenitani2018}, which reduces the phase errors of the original algorithm while keeping its desirable phase space conservation properties \cite{Qin2013}.

To test the prediction in Eq.~(\ref{eq:xPolDeltaVzBar}), we performed a parameter sweep across:
\begin{align}
	\bar{A}_0 &\in \{10^{-4},10^{-3},10^{-2}\}\\
	\bar{\omega} &\in \{0.01, 0.03, 0.1, 0.3, 0.5, 0.7, 0.9,\notag \\
	&\hspace{0.23in} 0.95, 0.97, 1.03, 1.05,1.1,1.3,1.5\}\\
	\bar{v}_{z 0} &\in \{-1,-0.5,0,0.5,1\}\\
	\bar{v}_{\perp 0} &\in \{0,1\}.
\end{align}
For each parameter set, 13 simulations were carried out, corresponding to different evenly-spaced initial gyro-angles.
In each simulation, the wave field was ramped up over the course of $\bar{\tau}_R = 10000$ inverse gyrofreqencies, and then held constant for a further 1000.
The reported $\Delta \bar{v}_z$ for each parameter set was then given by the average final velocity over the last 1000 inverse gyroperiods, further averaged over the 13 initial gyro-angles.
Because these simulations were relatively inexpensive on a modern laptop, we used a small timestep $\Delta \bar{t} = 0.07 \times \min (\bar{\alpha}^{-1},1)$.

The results for $\bar{v}_{\perp 0} = 0$ are shown in Fig.~\ref{fig:xPolSim}a, and for $\bar{v}_{\perp 0} = 1$ (i.e. $k_\parallel \rho = 1)$ in Fig.~\ref{fig:xPolSim}b, where they are compared to the theoretical prediction from Eq.~(\ref{eq:xPolDeltaVzBar}).
Two facts are immediately apparent.
First, the two cases are very different, with the sign of the force actually switching sign in the region $-1 < \bar{\alpha < 1}$ depending on the value of $\bar{v}_\perp$.
Second, in both cases, the agreement to simulations is quite good, except near the cyclotron resonances where the nonresonant approximation breaks down.
Thus, it is clear that the equation holds not only for the Doppler-shifted cold fluid ($\bar{v}_\perp = 0$), but also for the fully kinetic plasma ($\bar{v}_\perp = 1$) as well.



\section{Conclusion}

In this paper, we generalized the electrostatic theory of the nonresonant ponderomotive recoil to electromagnetic waves, demonstrating momentum conservation between resonant particles, nonresonant particles, and the electromagnetic fields.
Through single particle simulations, we verified that this form of the force holds even for general kinetic plasmas, 
By casting the recoil as the time derivative of a nonresonant momentum, we showed that the generalized Minkowski momentum of geometrical optics could be interpreted as the momentum gained by the nonresonant particles and electromagnetic fields as the wave self-consistently ramps up from vanishing amplitude.

The presence or absence of the recoil effect has implications both for magnetic field generation (magnetogenesis) and rotation drive effects in plasma.  
For laboratory magnetogenesis, the effects of the recoil will be most pronounced for direct current drive schemes such as LH current drive \cite{fisch1978currentDrive}, where the wave pushes resonant particles parallel to the background magnetic fields, including situations in which these waves might be amplified through the alpha channeling effect \cite{fisch1992current,ochs2015coupling,ochs2015alpha}.
However, they could also lead to consequences for indirect current drive schemes that rely on perpendicular electron heating, such as electron cyclotron current drive \cite{fisch1980creating}.  
In these systems, nonresonant effects are also likely to be important for the study of flows both parallel and perpendicular to the magnetic field, which often arise spontaneously \cite{Stoltzfus-Dueck2019} and which can stabilize instabilities and suppress turbulence across a variety of devices \cite{Shumlak1995,Zhang2019,Angus2020,Huang2001,Ellis2005,Cho2005,Ghosh2006,maggs2007transition,Gueroult2019,Burrell2020,Zweben2021}.
In addition to laboratory systems, the results could also be relevant in astrophysical settings, affecting magnetogenesis processes that occur through Landau damping \cite{Ochs2020a}, the radiative transfer dynamo effect \cite{munirov2019CMB}, or the inverse Bremstrahlung current drive effect \cite{Munirov2017InverseBremsstrahlung}.    
While in many of the above cases the application of interest is often steady state, in cases of start-up or transient situations, such as those driven by growing instabilities, the non-resonant effects addressed here can often play a large role.

\section*{Acknowledgments}

This work was supported by NNSA grant DE-SC0021248. 
This work was also supported by the DOE Fusion Energy Sciences Postdoctoral Research Program, administered by the Oak Ridge Institute for Science and Education (ORISE) and managed by Oak Ridge Associated Universities (ORAU) under DOE contract No. DE-SC0014664.

\section*{Data Availability}

Data sharing is not applicable to this article as no new data were created or analyzed in this study. 

\appendix

\section{Relationship to Electrostatic Theory} \label{app:EMES}

In this appendix, we verify that the electromagnetic force from Eq.~(\ref{eq:FsFull}) agrees with the previously derived electrostatic force from Refs.~\cite{ochs2021nonresonant,Ochs2021WaveDriven}, given by:
\begin{align}
	\llangle \ve{F}_{Es} \rrangle &= \frac{k_r^2 \phi^* \phi}{4\pi} \biggl[\lp  \frac{\lp \kapv \cdot \kv_r \rp \kv_r}{k_r^2}  - \frac{\kapv}{2}  \rp D_{rs} \notag\\
	&\hspace{0.35in} + \frac{\kv_r}{2} \lp D_{is} + \omega_i \pa{D_{rs}}{\omega_r} + \kapv \cdot \pa{D_{rs}}{\kv_r} \rp \biggr]. \label{eq:FEs}
\end{align}
Here, $\phi$ is the amplitude of the scalar potential field, and \begin{align}
	D_s \equiv -\frac{4\pi}{k^2} \frac{\rho_s}{\phi}
\end{align}
is the contribution of species $s$ to the (Poisson) dispersion relation. 
Furthermore, $D_{rs}$ and $D_{is}$ are the real and imaginary components of $D_s$ at real $\omega$ and $\ve{k}$, similarly to how $\chiv_s^H$ and $\chiv_s^A$ are the Hermitian and anti-Hermitian components of $\chiv_s$ at real $\omega$ and $\ve{k}$.

By using Eqs.~(\ref{eq:rho}) and (\ref{eq:j}), we can rewrite $D_s$ in terms of $\chi_s$:
\begin{align}
	D_s &= -\frac{4\pi}{k^2} \frac{1}{\phi} \lp -\frac{i}{4\pi} \kv \cdot \chiv_s \cdot \Ev \rp \\
	&= -\frac{1}{k^2} \frac{1}{\phi}  \lp -i \kv \cdot \chiv_s \cdot (-i \ve{k} \phi) \rp\\
	&= \frac{\kv \cdot \chiv_s \cdot \kv}{k^2},
\end{align}
which implies for $D_{rs}$ and $D_{is}$:
\begin{align}
	D_{rs} &= \frac{\kv_r \cdot \chiv^H_s \cdot \kv_r}{k_r^2}\\
	D_{is} &=\frac{\kv_r \cdot \chiv^A_s \cdot \kv_r}{k_r^2}.
\end{align}

Now we can just plug these substitutions into the electrostatic force in Eq.~(\ref{eq:FEs}).
The only complicated term is the last term in parentheses, which is best tackled in summation notation:
\begin{align}
	\kapv \cdot \pa{D_{rs}}{\kv_r} &= \kappa^\lambda \pa{}{k_r^\lambda} \lp \frac{k_r^i \chi_{s\, ij}^H k_r^j}{k_{rl} k_r^l} \rp \\
	&= \frac{2}{k_r^4} \kappa^\lambda k_\lambda k_r^i \chi_{s\, ij}^H k_r^j + \frac{2}{k_r^2} \Re \lp \kappa^\lambda \chi_{s \lambda j}^H k_r^j \rp \notag\\
	&\quad + \frac{1}{k_r^2} \kappa_\lambda k_r^i \pa{\chi_{s ij}^H}{k_r} k_r^j \\
	&= \frac{2}{k_r^4} \lp \kapv \cdot \kv_r\rp \kv_r \cdot \chiv_{s}^H \cdot \kv_r + \frac{2}{k_r^2} \Re \lp \kapv \cdot \chiv_s^H \cdot \kv_r \rp \notag\\
	&\quad + \frac{1}{k_r^2}\kv_r \cdot \left[\lp \kapv \cdot \pa{}{\kv_r}\rp\chiv_s^H \right] \cdot \kv_r .
\end{align}
After a bit more algebra, and recognizing $\ve{E} = -i\ve{k}_r \phi$, the electrostatic force becomes: 
\begin{align}
	\llangle \ve{F}_{Es}\rrangle &= \frac{1}{4\pi} \Re \biggl\{ \ve{E}^* \lp \kapv \cdot \chiv_s^H \cdot \ve{E} \rp    - \frac{1}{2} \kapv \lp \Ev^* \cdot \chiv_s^H \cdot \Ev \rp \notag\\
	&\hspace{0.1in} + \frac{1}{2}\ve{k}_r \lp \ve{E}^* \cdot \lp \chiv_s^A + \omega_i \pa{\chiv_s^H}{\omega_r} + \lp \kapv \cdot \pa{}{\ve{k}_r} \rp \chiv_s^H \rp  \cdot \ve{E} \rp \biggr\}. 
\end{align}
We can recognize this expression as the full electromagnetic force in Eq.~(\ref{eq:FsFull}), except without the second line, which goes to 0 for an electrostatic wave (with $\nabla \times \ve{E} = 0$).
Thus, the electromagnetic ponderomotive force derived here is consistent with the simpler form derived earlier for electrostatic waves \cite{ochs2021nonresonant,Ochs2021WaveDriven}.

\section{Consistency with Maxwell Stress Tensor} \label{app:MaxwellStress}

Here, we show that the ponderomotive force on the plasma in Eq.~(\ref{eq:FsFull}) in the presence of spatial gradients is consistent with force exerted by the electromagnetic fields on the plasma through the Maxwell stress tensor.
The (negative) Maxwell stress tensor is given by:
\begin{align}
	\Piv_{EM} &= \frac{1}{4\pi} \lp - \Ev \Ev + \frac{1}{2} \ve{I} E^2 - \Bv \Bv + \frac{1}{2} \ve{I} B^2 \rp,
\end{align}
from which the total force on the medium is:
\begin{align}
	\llangle \ve{F}_{EM} \rrangle &= -\llangle \nabla \cdot \Piv_M \rrangle\\
	&= \frac{1}{4\pi} \Re \biggl[\ve{\kapv} \cdot \biggl( - \Ev \Ev^* + \frac{1}{2} \ve{I} \lp \Ev \cdot \Ev^*\rp  \notag\\
	&\hspace{0.85in} - \Bv \Bv^* + \frac{1}{2} \ve{I} \lp \Bv \cdot \Bv^*\rp \biggr) \biggr].
\end{align}

Now, consider the ponderomotive force from Eq.~(\ref{eq:FsFull}) with $\omega_i = 0$.
Summing this force over all species in the plasma, we can write this force as:
\begin{align}
	\sum_s \llangle \ve{F}_s \rrangle &= \frac{1}{4\pi} \Re \biggl\{ \ve{X} + \ve{Y} + \ve{Z} \biggr\},
\end{align}
where
\begin{align}
	\ve{X} &\equiv \sum_s \ve{E}^* \lp \kapv \cdot \chiv_s^H \cdot \ve{E} \rp \label{eq:X1}\\
	\ve{Y} &\equiv \sum_s -\frac{1}{2} \kapv \lp \Ev^* \cdot \chiv_s^H \cdot \Ev \rp \\
	\ve{Z} &\equiv \sum_s \frac{1}{2}\ve{k}_r \lp \ve{E}^* \cdot \lp \chiv_s \rp_A  \cdot \ve{E} \rp.
\end{align}

The latter two of these terms are straightforward to simplify. 
We have:
\begin{align}
	\Re \, \ve{Y} &= -\frac{1}{2} \kapv \Re \left[ \Ev^* \cdot \lp -\Ev - \ve{n}_r \times \lp \ve{n}_r \times \Ev \rp \rp \right]\\
	&= \frac{1}{2} \kapv \Re \left[ \Ev^* \cdot \Ev - \lp \ve{n}_r \times \Ev^* \rp \cdot \lp \ve{n}_r \times \Ev \rp \right]\\
	&= \frac{1}{2} \kapv \Re \left[ \Ev^* \cdot \Ev - \Bv^* \cdot \Bv \right],
\end{align}
and
\begin{align}
	\Re \, \ve{Z} &= -\frac{\kv_r}{2}\frac{c^2}{\omega_r^2} \Re \biggl[\ve{E}^* \cdot \biggl( \kapv \times \lp \ve{k}_r \times \ve{E} \rp \notag\\
	& \hspace{0.65in} + \kv_r \times \lp \kapv_r \times \ve{E} \rp \biggl) \biggr]\\
	&= \frac{\kv_r}{2}\frac{c^2}{\omega_r^2}  \Re \biggl( \lp  \kapv \times  \ve{E}^* \rp \cdot \lp \ve{k}_r \times \ve{E} \rp \notag\\
	& \hspace{0.5in} + \lp  \kv \times  \ve{E}^* \rp \cdot \lp \kapv \times \ve{E} \rp \biggl)\\
	&= \ve{n}_r \Re \left[\lp \kapv \times \Ev\rp \cdot \Bv^* \right].
\end{align}
In simplifying $\ve{Z}$, we used the fact that $\ve{a} \times \lp \ve{a} \times \ve{E} \rp$ is Hermitian when $\ve{a}$ is real, as well as the Taylor expansion:
\begin{align}
	\ve{n}_i &= \Im \lp \frac{(\ve{k}_r + i \kapv) c}{\omega_r} \rp = \frac{\kapv c}{\omega_r}.
\end{align}

Now, define the quantity $\ve{A}$ as follows:
\begin{align}
	\ve{A} &\equiv 4\pi \lp \llangle \ve{F}_{EM} \rrangle  - \sum_s \llangle \ve{F}_s \rrangle \rp  + \Re \, \ve{X}\\
	&= \Re \biggl[- \Ev^* \lp \kapv \cdot \Ev \rp \underbrace{- \Bv^* \lp \kapv \cdot \Bv \rp +  \kapv \lp \Bv^* \cdot \Bv \rp}_\ve{C} \notag\\
	&\hspace{0.8in} - \ve{n}_r \lp \kapv \times \Ev \cdot \Bv^* \rp \biggr]. \label{eq:A1}
\end{align}
If we show that $\ve{A} = \Re \, \ve{X}$, then $\llangle \ve{F}_{EM} \rrangle = \sum_s \llangle \ve{F}_s \rrangle $.

To show this, we begin by simplifying the quantity $\ve{C}$ in Eq.~(\ref{eq:A1}):
\begin{align}
	\ve{C} &= \Bv \times \lp \kapv \times \Bv^* \rp\\
	&= \lp \ve{n}_r \times \Ev \rp \times \lp \kapv \times \Bv^*  \rp\\
	&= - \lp \kapv \times \Bv^*  \rp \times \lp \ve{n}_r \times \Ev \rp\\
	&= \lp \kapv \times \Bv^* \cdot \ve{n}_r \rp \Ev - \lp \kapv \times \Bv^* \cdot \Ev \rp \ve{n}_r\\
	&= -\lp \kapv \cdot \ve{n}_r \times \Bv^* \rp \Ev + \lp \kapv \times \Ev \cdot \Bv^* \rp \ve{n}_r.
\end{align}
Pluging this back into Eq.~(\ref{eq:A1}), several terms cancel, and we have:
\begin{align}
	\ve{A} &= \Re \left[ - \Ev^* \lp \kapv \cdot \Ev \rp -\Ev \lp \kapv \cdot \ve{n}_r \times \lp \ve{n}_r \times \Ev^* \rp \rp   \right]\\
	&= \Re \left[ \Ev^* \lp \kapv \cdot \lp - \Ev - \ve{n}_r \times \lp \ve{n}_r \times \Ev \rp \rp \rp   \right]
\end{align}
Now, we can plug in our 0th order dispersion relation (Eq.~(\ref{eq:DispZerothOrder})), yielding:
\begin{align}
	\ve{A} &= \Re \left[ \Ev^* \lp \kapv \cdot \lp \sum_s \chiv_s \cdot \Ev  \rp \rp   \right]
\end{align}
Comparing to Eq.~(\ref{eq:X1}), we see that the quantity in brackets is just $\ve{X}$.
Thus, $\ve{A} = \Re \, \ve{X}$, and so $\llangle \ve{F}_M \rrangle  = \llangle \ve{F} \rrangle$, meaning that the ponderomotive force is consistent with the total force on the plasma from the electromagnetic stress.

\input{EMRecoilResub.bbl}

\clearpage

\end{document}

%% file: EMRecoilResub.bbl
%